\documentclass[
aps,
prl,
showpacs,
amsmath,
amssymb,
reprint,
floatfix,
lengthcheck,
superscriptaddress,
groupedaddress
]{revtex4-1}

\usepackage{graphicx}
\usepackage{mathtools}
\usepackage{braket}
\usepackage{hyperref}
\usepackage{color}
\usepackage[normalem]{ulem}
\usepackage{algorithmicx}
\usepackage{algorithm}
\usepackage{algpseudocode}

\makeatletter \renewcommand\@make@capt@title[2]{%
\@ifx@empty\float@link{\@firstofone}{\expandafter\href\expandafter{\float@link}}%
\sffamily{\textbf{#1}}\@caption@fignum@sep#2 }

\newcommand{\HarvardSEAS}{John A. Paulson School of Engineering and Applied Sciences, Harvard University, Cambridge, MA, USA}

\begin{document}

\title{Qubit Allocation for Noisy Intermediate-Scale Quantum Computers}

\author{Will Finigan}
\thanks{These two authors contributed equally}
\affiliation{\HarvardSEAS}

\author{Michael Cubeddu}
\thanks{These two authors contributed equally}
\affiliation{\HarvardSEAS}

\author{Thomas Lively}
\altaffiliation{Present address: Google LLC, 1600 Amphitheatre Parkway, Mountain View, CA 94043}
\affiliation{\HarvardSEAS}

\author{Johannes Flick}
\email{flick@seas.harvard.edu}
\affiliation{\HarvardSEAS}

\author{Prineha Narang}
\email{prineha@seas.harvard.edu}
\affiliation{\HarvardSEAS}

\date{\today}

\begin{abstract}
In the era of noisy-intermediate-scale quantum computers, we expect to see quantum devices with increasing numbers of qubits emerge in the foreseeable future. To practically run quantum programs, logical qubits have to be mapped to the physical qubits by a qubit allocation algorithm. However, on present day devices, qubits differ by their error rate and connectivity. Here, we establish and demonstrate on current experimental devices a new allocation algorithm that combines the simulated annealing method with local search of the solution space using Dijkstra's algorithm. Our algorithm takes into account the weighted connectivity constraints of both the quantum hardware and the quantum program being compiled. New quantum programs will enable unprecedented developments in physics, chemistry, and materials science and our work offers an important new pathway toward optimizing compilers for quantum programs.
\end{abstract}

\date{\today}

\maketitle

+It is the general belief that quantum information science will enable unprecedented developments in many fields of research, including physics, chemistry~\cite{mcArdle2018}, biology~\cite{reiher2017} and materials sciences~\cite{king2018}.
The existing quantum technology, recently termed ``noisy-intermediate scale quantum'' (NISQ) technology~\cite{preskill2018}, has already allowed researchers to realize systems with fewer than 100 qubits in experimental laboratories~\cite{castelvecchi,kandala2017,zeng2017} and inspired many theoretical developments~\cite{hormozi2007,takita2016,takita2017,dahlberg2019,ding2018,yunger2018,bouland2018, dalzell2018}.

One of the current challenges of quantum computing is that quantum computations are noisy and have nonzero rates of error. These error rates are due to multiple factors: qubits can only stay in a mixed state for a certain period of time, the entangling and rotational operations are susceptible to accuracy errors, and subtle changes in the surrounding environment can affect qubit operation accuracy. Reducing the error rates of the executions is essential for maximizing the reliability of the results produced by NISQ computers.

One way to efficently reduce noise for quantum computations is by introducing a pre-processing step to optimize or compile the quantum circuit to the specifications of the underlying hardware. One important optimization step within the compiler is \textit{qubit allocation}~\cite{siraichi2018}. The allocation of qubits is the process of mapping logical qubits in the quantum program to physical qubits on the hardware. Qubit allocation must satisfy the connectivity constraints of both the program and the hardware while minimizing the total error rate of the computation.

There are several proposed methods for solving the qubit allocation problem. Recent studies~\cite{zulehner2018, siraichi2018,shafaei2013, li2018} have proposed minimizing the number of \textsc{SWAP} instructions for efficient qubit allocation. The introduction of \textsc{SWAP} gates are required on NISQ devices due to the limited connectivity of the actual device. However, additional \textsc{SWAP} gates increase the number of operations in the circuit, as well as the depth of the circuit. While the number of \textsc{SWAP}s in a circuit strongly affects the total error rate of the computation, it is not the only factor the compiler should consider when allocating qubits. As has been recently discussed in Ref.~\cite{tannu2018}, since in reality not all qubits are created equal, the assumption of a uniform behavior of the qubits is not justified. To this end, Ref.~\cite{tannu2018} introduced the variation-aware qubit allocation (VQA) policy that considers the different error rates of the individual qubits during allocation. VQA has been shown to improve the overall reliability of computation results significantly.


In this letter, we propose an algorithm that sets a new standard for compilers for NISQ-era computers.
The proposed algorithm employs the method of simulated annealing~\cite{kirkpatrick1983} to sample the vast search space of qubit allocation. We increase the reliability of the simulated annealing by complementing it with a bounded search of the local solution space using Dijkstra's algorithm ~\cite{knuth1977}. The proposed algorithm is \emph{hardware-agnostic}, since it can be parameterized with the particulars of any quantum computing architecture. The algorithm is also tunable, allowing users to trade additional compute time for better results. Additionally, we benchmark our algorithm and show that the solutions proposed in this letter significantly improve on existing methods. These benchmarks are one of the first times a proposed quantum compiler has been formally tested on physical hardware. Therefore, our proposed solution pushes the field towards a more practical solution to the qubit allocation problem.

\begin{figure}[h]
      \centering
      \includegraphics[width=0.45\textwidth]{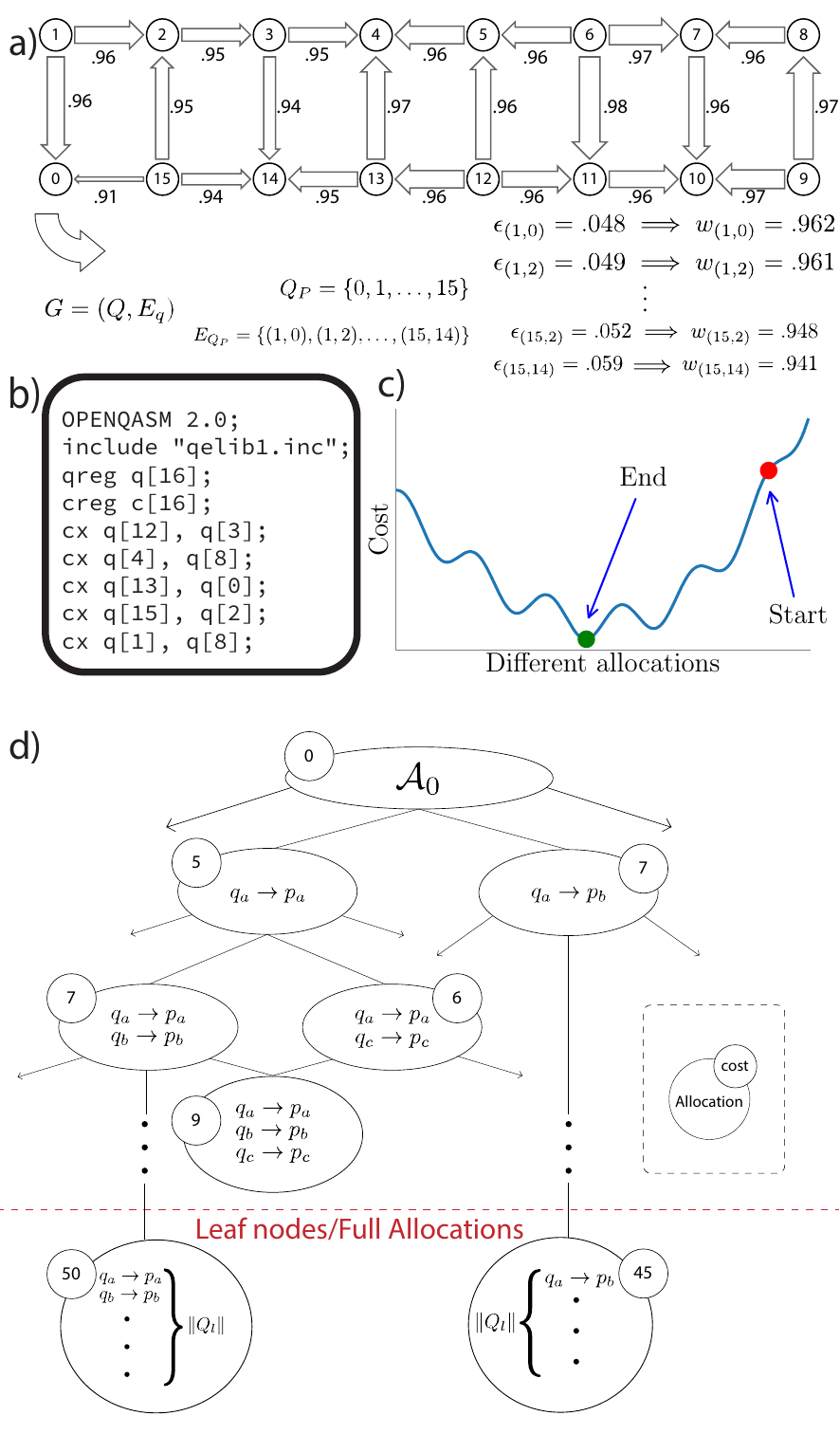}
      \caption{
      (a): Layout of IBM 16Q Rueschlikon device, showing variable fidelities and coupling graph. (b): Example OpenQASM program, consiting of randomly generated CNOTS. These programs will be discussed in more detail in the benchmark section. (c) Schematic describing the search space of the annealing algorithm discussed in this paper. Importantly we notice the presense of local minima. (d) Representation of the DAG, $G_{P, D}$ as constructed in the local search algorithm.
      }
      \label{fig:define_problem}
 \end{figure}

In this letter, we consider quantum programs written in the Quil~\cite{quil} and OpenQASM~\cite{qasm} languages. We show an example of a quantum program in Fig.~\ref{fig:define_problem} (b). For a given quantum program $P$, let an \emph{allocation} $\mathcal{A}$ be a mapping of a set $Q_L$ of logical qubits in $P$ to distinct physical qubits $Q_P$ of a device.
\begin{align}
\mathcal{A}: Q_L \underset{P}{\rightarrow} Q_P.
\end{align}
Further, we define as the size of an allocation $|\mathcal{A}| = |Q_P|$, which is the number of allocated qubits. The empty allocation $\mathcal{A}_0$ is the allocation for which $|\mathcal{A}_0| = 0$, and a \emph{full allocation} $\mathcal{A}_F$ for $P$ is an allocation  where all logical qubits are mapped to physical qubits $Q_P$. In addition, we define a \emph{sub-allocation} relation, $\sqsubset$, such that for allocations $\mathcal{A}_A$ and $\mathcal{A}_B$, $\mathcal{A}_A \sqsubset \mathcal{A}_B$ if every mapping in $\mathcal{A}_A$ is also present in $\mathcal{A}_B$.

As an abstract representation of different hardware devices, we choose to represent different systems by a \emph{coupling graph}~\cite{siraichi2018} $G_{Q_P}=(Q_P,E_{Q_P})$, where $Q_P$ is the set of physical qubits on the hardware, and $E_{Q_P}$ is the set of edges connecting the physical qubits. We illustrate the coupling graph of a 16-qubit system in Fig.~\ref{fig:define_problem} (a). All edges $e\in E_{Q_P}$ have an associated error rate $\epsilon_e$. The error rate $\epsilon_e$ of an edge $e$ ranges from 0 to 1. Due to the imperfection of the NISQ devices, there is variation in the error rates of different qubits and edges over time. These error rates are measured during device calibrations and occasionally published~\cite{tannu2018}. We further define the fidelity or reliability of an edge as $\mathcal{F}_e=1-\epsilon_e$ and define as $\mathcal{F}^{(i)}$ the fidelity of a $i$-qubit gate when applied to the physical qubits.

Since most existing quantum computers use superconducting qubits~\cite{castelvecchi, kandala2017,zeng2017}, these systems are only able to execute two-qubit gates on immediately connected qubits as depicted in Fig.~\ref{fig:define_problem} (a). Therefore, the highest connectivity is usually two and we can restrict ourselves to $i=\{1,2\}$.

Further, we define the total fidelity of the computation as the product of all fidelities of $N_1$ single-qubit gates and $N_2$ two-qubit gates that are applied on the physical hardware devices during runtime of the quantum program
\begin{align}
\mathcal{F}_\text{tot} = \prod_{i=1}^{N_2} \mathcal{F}^{(2)}_{i} \times \prod_{j=1}^{N_1}\mathcal{F}^{(1)}_{j}.
\label{eq:etot}
\end{align}
In general two-qubit error rates are one magnitude higher than single qubit error rates, and the total error rate $\mathcal{E}_\text{tot}$ is therefore dominated by the two-qubit error rates. In this letter, we develop an algorithm that maximizes the total fidelity for hardware specific connectivity constraints by minimizing the total error rate. In Fig.~\ref{fig:define_problem} (c), we illustrate the complex fidelity landscape of the qubit allocation problem schematically.

To resolve situations where two logically adjacent qubits are not physically connected on the specific hardware, \textsc{SWAP} gates have to be inserted into the quantum program~\cite{siraichi2018, tannu2018}, which introduces additional computational overhead. To compute the optimal \textsc{SWAP} paths we calculate the optimal \textsc{SWAP} path for every ordered pair of two physical qubits on the hardware as a pre-processing step in $\mathcal{O}(|Q_P|^3)$ time using the Floyd-Warshall algorithm~\cite{warshall1962}. This is similar to the analysis done in Ref.~\cite{tannu2018}, which uses Dijkstra's algorithm instead. For our algorithm, we simplify the qubit allocation problem by considering \textsc{SWAP} placement only in cases where swaps are necessary to satisfy the connectivity constraints of the quantum devices, and when they are necessary we insert them as late as possible. In the following, we call this policy \emph{connectivity-only swap insertion}~\footnote{In the general qubit allocation problem, optimal solutions may contain swaps that are not necessary to satisfy the constraints of the problem, but are able to reduce the error rate of the quantum program.}. This simplification makes the set of necessary swaps necessary for a starting qubit allocation computable in time linear with the size of the program.

The problem of qubit allocation with connectivity-only \textsc{SWAP} insertion can be reduced to the problem of finding the shortest weighted path in a directed acyclic graph (DAG)~\cite{thulasiraman1992}, $G_{D,P}$, which we illustrate in Fig.~\ref{fig:define_problem} (d). The set of vertices of $G_{P,D}$ is the set of all possible allocations, including partial allocations, for the quantum program $P$ and device $D$. Its set of edges is the set of all pairs of allocations $(\mathcal{A}, \mathcal{A^\prime})$ with $|\mathcal{A}| + 1 = |\mathcal{A^\prime}|$ and $\mathcal{A} \sqsubset \mathcal{A^\prime}$. Fig.~\ref{fig:define_problem} (d) illustrates several important properties of the graph $G_{P,D}$. First, the root of $G_{P,D}$ is the empty allocation, and every vertex is a sub-allocation of all other vertices reachable from it. Edges represent the extension of allocations by a single additional qubit mapping. Furthermore, $G_{P,D}$ is acyclic because every vertex in $G_{P,D}$ is strictly larger than each of its predecessors, and a vertex is a leaf if and only if it is a full allocation.

Each path from $\mathcal{A}_0$ to a full allocation $\mathcal{A}_F$ represents a sequence of decisions to allocate single qubits that together form a full allocation, so $G_{P,D}$ represents every possible way to create a full allocation starting from $\mathcal{A}_0$. To find solutions to the qubit allocation problem, we need to find paths that connect $\mathcal{A}_0$ to some $\mathcal{A}_F$. By appropriately constructing the edge weights in $G_{P,D}$, we can ensure that the weighted shortest paths end at the full allocation with the highest fidelity.

Following Eq.~\ref{eq:etot}, we define the upper bound $\mathcal{F}_A$ for the allocation $\mathcal{A}$ to be the total fidelity calculated using the mappings in $\mathcal{A}$, including any swaps they necessitate, and by taking for all unallocated logical qubits the best possible physical qubits without considering connectivity or uniqueness constraints. We notice that when the allocation is full, this bound is the true fidelity $\mathcal{F}_\text{tot}$. Now let the weight of edge $(\mathcal{A}, \mathcal{A^\prime})$ be $\mathcal{F}_A - \mathcal{F}_{A^\prime}$. By the definition of edges in $G_{P,D}$, $\mathcal{A^\prime}$ contains every qubit allocation in $\mathcal{A}$ plus one more. The the additional allocated qubit tightens the calculated upper bound on the fidelity of gates, so $\mathcal{F}_{A^\prime}$ must be less or equal to $\mathcal{F}_A$ and the edge weight cannot be negative.
More generally, the weight of a path from $\mathcal{S}$ to $\mathcal{T}$ is the sum of the weights of all edges in that path, which collapses to $\mathcal{F}_S-\mathcal{F}_T$. The problem of optimal qubit allocation with connectivity-only swap insertion is thereby reduced to the problem of finding the shortest path in a weighted DAG with positive edge weights. This problem can be solved efficiently by Dijkstra's algorithm~\cite{knuth1977}.

Note that our qubit allocator is also able to handle arbitrary classical control flow~\cite{bohm1966}. The Quil language~\cite{quil} provides \textsc{JUMP}, \textsc{JUMP-IF}, and \textsc{JUMP-UNLESS} instructions that can transfer control flow elsewhere in the program. Nevertheless, all considerations of control flow are encapsulated in the calculations of the swap set and fidelity bound at each step of the local search. The higher level qubit allocation algorithms are therefore agnostic to the presence or absence of control flow. The details of how we handle control flow will be published elsewhere.

As an optimization, the number of paths searched in the DAG can be reduced by fixing the order in which qubits are allocated, turning the $G_{D,P}$ into a tree $G_{D,P}^\prime$~\footnote{Which order the qubits are allocated in does not matter for accuracy, but may matter for execution time.}.
Our implementation heuristically chooses to allocate logical qubits from most to least constrained. We define a logical qubit to be more constrained than another if it appears as the control in more two-qubit gates than the other. By allocating the most constrained qubits first, we minimize the chance of an allocation we have spent a lot of time analyzing becoming untenable due to unsatisfiable constraints.

In the worst case, when all possible allocations have the same fidelity, Dijkstra's algorithm is equivalent to a breadth-first search~\cite{lee1961}. For a device with $Q_P$ physical qubits, the number of possible allocations of size $n$ is $\frac{Q_P!}{(Q_P- n)!}$, thus the number of edges traversed in the worst case for a program with $Q_L$ logical qubits is
$\sum_{n = 1}^{Q_L}\frac{Q_P!}{(Q_P - n)!}$.
Current devices are sparsely connected and have non-uniform qubit fidelities, so they are far from exhibiting these worst-case properties. Djikstra's algorithm will not consider any vertex whose fidelity bound is lower than the true optimal fidelity, so in practice the algorithm might only visit a very small portion of the solution space before finding the optimal solution. For all allocations to have the same fidelity, either a device's qubit connectivity graph must be complete and all of its couplings must have the same fidelities, or the connectivity graph of the logical qubits must be symmetric, for example if each logical qubit participates in a gate with each other logical qubit exactly once and there are no additional gates.

Unfortunately, the existence of quantum programs that can trigger the exponential behavior of the local search algorithm make it unfit to serve as a complete practical solution. Nevertheless, this algorithm is very useful as a component of such a solution.

\begin{figure}[h]
      \centering
      \includegraphics[width=0.55\textwidth]{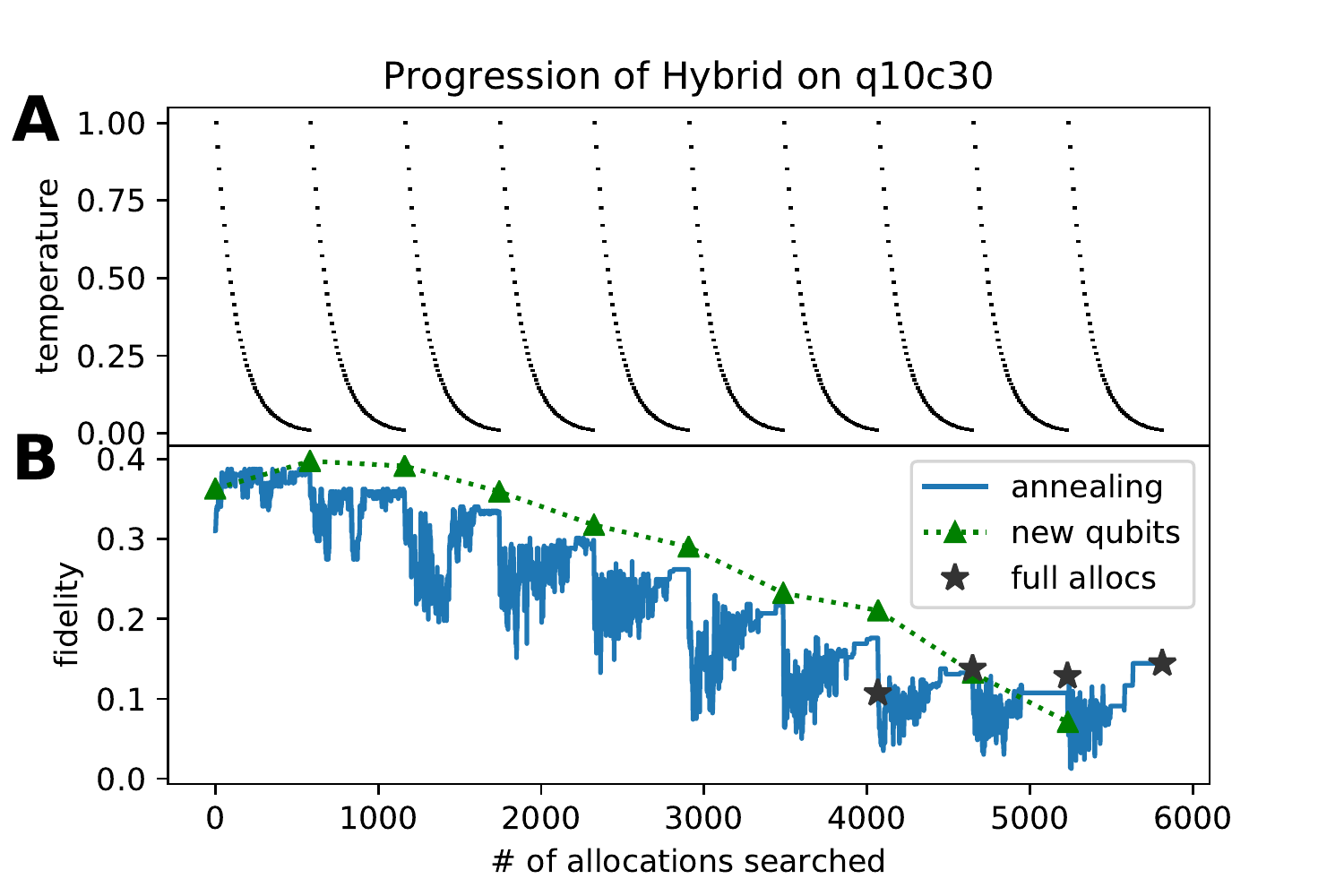}
      \caption{Hybrid algorithm on a randomly generated 10-qubit, 30 CNOT Quil~\cite{quil} circuit called \texttt{q10c30} with ${n} = 5$, $T_0=1$, and $\tau=12.5$. \textbf{2.A} shows the change in the temperature $T$ as the simulated annealing routine explores allocations. The annealing data in graph \textbf{2.B} shows the fidelity bounds of all the partial allocations the hybrid algorithm explores, adding one qubit for each round of annealing. Full allocations are denoted by a star.}
  \label{fig:algorithm}
\end{figure}

To this end, we introduce randomization using simulated annealing~\cite{kirkpatrick1983} as a method of reducing the search space of the local search algorithm. In the following, we refer to this randomized optimization procedure combined with the local search algorithm as the \textit{hybrid algorithm}.

The hybrid algorithm finds a full allocation by running the simulated annealing process on sets of progressively larger partial allocations. For each proposed sub-allocation, we run the local search for a fixed number of steps to produce a refined estimate of the sub-allocation's fidelity. During each round of simulated annealing, we use the Metropolis criterium~\cite{chib1995} to accept sub-allocations: If a proposed sub-allocation has a higher fidelity bound than the current sub-allocation, we accept it as a better starting point for the local search. If it has a lower fidelity bound, we accept it with a probability dependent on the fidelity difference to prevent the search from getting trapped in a local minima as shown in Fig.~\ref{fig:define_problem} (c). The probability of accepting a sub-allocation with lower estimated fidelity also depends on the current temperature that we define as $T = T_0 \, \text{exp}(-s/\tau)$, where $s$ is the iteration counter within a simulated annealing process, $T_0$ the initial temperature, and the reduction of the temperature is given by $\tau$ as illustrated in Fig.~\ref{fig:algorithm} (a).


Initially, we run the local search algorithm for $n$ steps starting with the empty allocation to see if the local search can find the optimal allocation in reasonable time without the need for randomization.
If this search fails, we use simulated annealing to allocate a single logical qubit.
Simulated annealing proposes potential allocations for this qubit and uses local search to refine an upper bound on the cost of a full allocation containing the proposed sub-allocation.
The algorithm adds one qubit at a time to the annealing search as shown in Fig.~\ref{fig:algorithm} (b), where we show the fidelity bounds of the allocations the hybrid algorithm explores during a run on ten qubits. Each time local search is called to measure the cost, it is also searching for potential full allocations based on the proposed sub-allocation. Before adding another qubit to the simulated annealing search, we check whether any full allocations have been found.

If local search finds at least one full allocation during a given round of simulated annealing, we return the full allocation with the highest fidelity. For illustrative purposes, we show a full run in Fig.~\ref{fig:algorithm} (b). When the temperature $T$ is higher, we are more likely to accept worse proposed allocations and when it is lower we tend to only accept better or similar allocations. The idea motivating this design is that as a simulated annealing round progresses, we want to explore the solution space less and instead refine the best solution found so far. This explains why there is a reduction in noise throughout each of the $10$ annealing iterations in Fig.~\ref{fig:algorithm} (b) as the respective temperatures plotted in Fig.~\ref{fig:algorithm} (a) decrease. Generally, we expect to see a downward trend in fidelity bounds as new qubits are added to the search, as shown in Fig.~\ref{fig:algorithm} (b), because larger allocations have tighter upper bounds on total fidelity than smaller allocations. We highlight that the simulated annealing process is easily parallelizable and can be run several times simultaneously.

Next, we compare our algorithms, both local search and the hybrid algorithm, to other methods in the field. We focus on testing against publicly available tools provided by IBM Q Experience by running randomly generated CNOT circuits the Q16 Rueschlikon device. As there is no standard method for testing total circuit fidelity, we propose a new benchmark solution. Our testing framework is as follows: first we create a randomly generated circuit containing only CNOT gates. Next we compile this circuit into a device executable OpenQASM file through three methods: IBM's public compiler, our local search allocator, and our hybrid algorithm with parameters $n=10$, $T_0=10$, and $\tau=25$. Next, we run each of these compiled algorithms on IBM's hardware measuring one qubit at a time. We do this to avoid noise associated with measuring multiple qubits successively.

Since each circuit contains only CNOT gates, in principle the qubit measured should have 100\% probability of being measured in the ground state.
Therefore, if a qubit is measured in the excited state, we know this must be a product of device noise.
So, we define the measured error of a circuit to be the number of total number of trials in which each qubit involved in the circuit is measured to be in the excited state. The percent error is the percentage of incorrect measurements in all trials of a circuit.
To minimize noise associated with single qubit measurement, we run the circuit 1024 times for each qubit involved in the circuit, measuring the physical qubits one at a time.
\begin{figure}[H]
    \centering
    \includegraphics[width=0.5\textwidth]{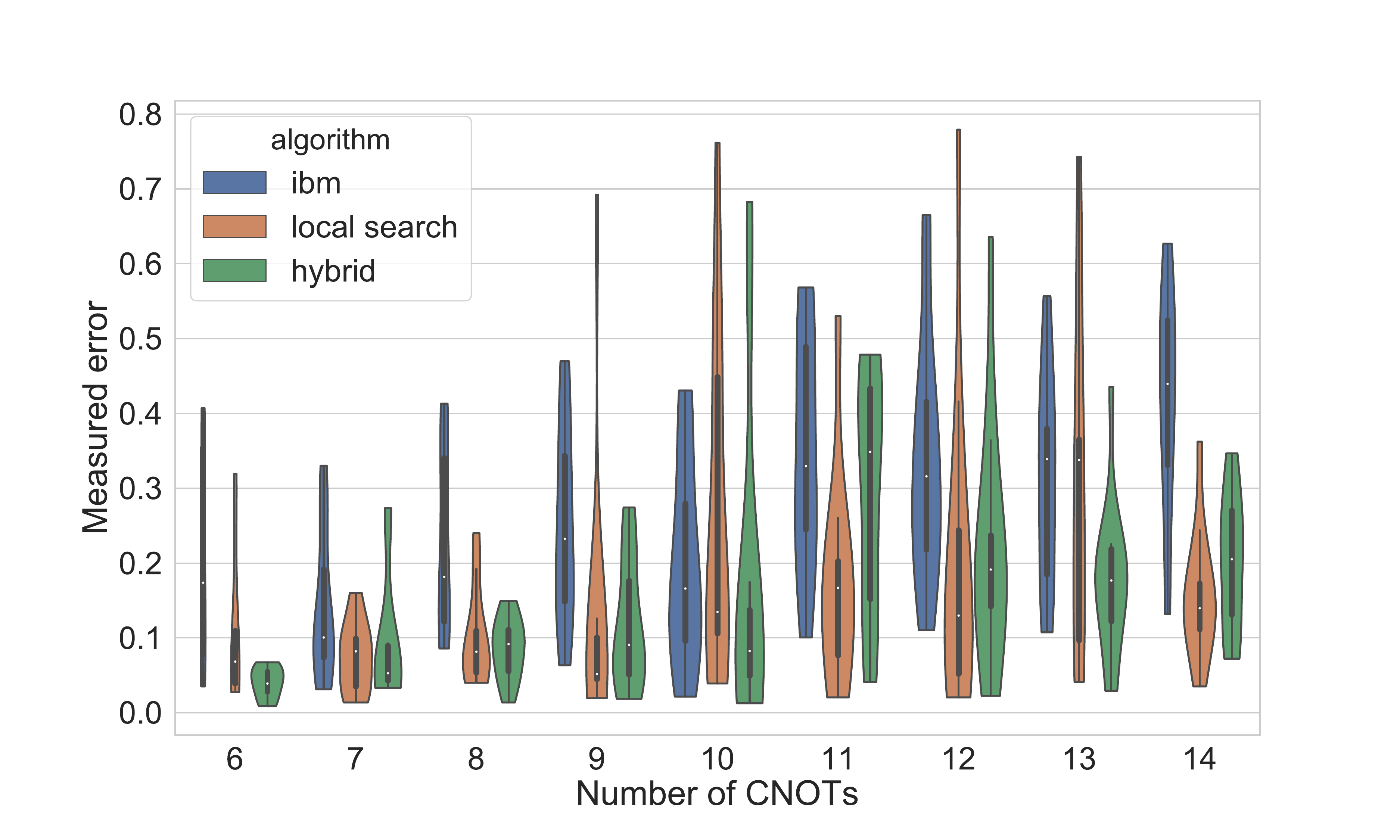}
    \caption{Measured error by compiling a series of randomly generated CNOT circuits using each method successively and running IBM's Q 16 device.}
  \label{fig:results}
\end{figure}
The violin plot in in Fig.~\ref{fig:results} shows the distribution of the measured error for each circuit compared by method, where a single observation is computed by counting the number of measurements in the excited state and dividing by the total number trials (1024).
The width of the plot corresponds to the number of observations measured at each value.
Within each distribution is a standard box plot, showing the median (denoted by a white dot) as well as the first and and third quartiles.
Our results presented in Fig.~\ref{fig:results} show that the local search method consistently has a lower error rate than IBM's public compiler, resulting in more accurate measurements in every case.
This advantage tends to increase as the size of the circuit increases.
The hybrid algorithm tends in general to show more accurate results compared to the IBM's compiler as well, although with more variability.
All three algorithms resulted in a high variability in measurement error.
Some of these error rates (greater than 0.7 in certain cases) suggest imprecision in the listed measurement errors.
These outliers seem to be the result of device peculiarities, rather than the known errors referred to in this paper.
These are exciting results for the potential of NISQ-era devices, as they demonstrate that we can reduce device noise by over a factor of 2 using pre-processing.

NISQ-era computers will change the landscape of technology through their capability of solving complex computer science problems that classical computers are unable to solve efficiently~\cite{reiher2017}. We believe that the qubit allocation algorithms we propose in this paper can be used for compilation on these intermediate-scale devices. As sub-problems of qubit the allocation problem are NP-complete, such as the \textsc{SWAP} minimization problem~\cite{siraichi2018}, the hybrid algorithm uses a randomized search to efficiently search the solution space while still obtaining a total fidelity close to that of the optimal solution. The parameters of the hybrid algorithm, $n$, $T_0$, and $\tau$ allow users to trade off between the time spent on allocation and total fidelity. The hybrid algorithm generalizes both a pure simulated annealing approach when $n = 0$ and the local search algorithm when $n\rightarrow\infty$.

Currently, there is ongoing debate and active research on differing physical implementations of quantum computing architectures~\cite{castelvecchi} (i.e. superconductors vs. ion traps), which each have different respective connectivity constraints between qubits.
The hybrid algorithm proposed in this paper is hardware-agnostic in that it works for any given coupling graph and qubit fidelity data, but it also takes full advantage of this hardware-specific information in its search for optimal qubit allocations.
We believe this dual nature of hardware-agnosticism and hardware-awareness provides a flexibility and sensitivity qualifying it as a practical tool for reducing the noise of quantum computation on NISQ computers.

We thank Antonio D. Corcoles-Gonzales for assistance on the IBM 16 Rueschlikon device and Yudong Cao for fruitful discussions. We further acknowledge support from the STC Center for Integrated Quantum Materials NSF grant number DMR-1231319 and from the Harvard John A. Paulson School of Engineering and Applied Sciences. J.F. acknowledges financial support from the Deutsche Forschungsgemeinschaft (DFG) under Contract No. FL 997/1-1. T.L. notes that this work is not associated with Google LLC or the Google Quantum AI Lab.

\bibliography{refs.bib} 

\end{document}